\documentclass[aps,prx,reprint,superscriptaddress]{revtex4-2}

\usepackage{graphicx}  

\usepackage{longtable}
\usepackage[T1]{fontenc}
\usepackage{dcolumn}   
\usepackage[inline]{enumitem}

\usepackage{bm}        
\usepackage{amsfonts}  
\usepackage{amsmath}   
\usepackage{amssymb}   

\usepackage{amsthm}
\usepackage{tikz}
\usetikzlibrary{quantikz}
\usepackage{glossaries}

\usepackage{color}
\usepackage[colorlinks=true,allcolors=blue]{hyperref}

\newcommand{\pwisein}{\left\{ \begin{array}{ll}}
\newcommand{\pwiseout}{\end{array}\right.}

\begin{document}

\title{Entanglement signature in quantum work statistics in the slow-driving regime}

\author{Jian Li}
\email{ji6174li-s@student.lu.se}
\affiliation {Department of Physics, Lund University, Box 118, 22100 Lund, Sweden}

\author{Mark T. Mitchison}
\email{mark.mitchison@tcd.ie}
\affiliation{School of Physics, Trinity College Dublin, College Green, Dublin 2, D02 K8N4, Ireland}
\affiliation{Trinity Quantum Alliance, Unit 16, Trinity Technology and Enterprise Centre, Pearse Street, Dublin 2, D02YN67, Ireland}

\author{Saulo V. Moreira}
\email{moreirsv@tcd.ie}
\affiliation {Department of Physics, Lund University, Box 118, 22100 Lund, Sweden}
\affiliation{School of Physics, Trinity College Dublin, College Green, Dublin 2, D02 K8N4, Ireland}

\begin{abstract}  
In slowly driven classical systems, work is a stochastic quantity and its probability distribution is known to satisfy the work fluctuation-dissipation relation, which states that the mean and variance of the dissipated work are linearly related.
Recently, it was shown that generation of quantum coherence in the instantaneous energy eigenbasis leads to a correction to this linear relation in the slow-driving regime.
Here, we go even further by investigating nonclassical features of work fluctuations in setups with more than one system. 
To do this, we first generalize slow control protocols to encompass multipartite systems, allowing for the generation of quantum correlations during the driving process.
Then, focussing on two-qubit systems, we show that entanglement generation leads to a positive contribution to the dissipated work, which is distinct from the quantum correction due to local coherence generation known from previous work. Our results show that entanglement generated during slow control protocols, e.g.~as an unavoidable consequence of qubit crosstalk, comes at the cost of increased dissipation.
 
\end{abstract}


\maketitle 

\section{Introduction}
The last few decades have seen the emergence of the fields of stochastic thermodynamics and quantum information. Classical stochastic thermodynamics entails fluctuations of thermodynamic quantities, such as work and heat, in non-equilibrium processes at the nanoscale. Examples of landmark research achievements include the discovery of fluctuation theorems~\cite{Bochkov1981, Bochkov1981b, Jarzynski2000, Crooks1998, Crooks2000}, which generalize the second law of thermodynamics for non-equilibrium systems, and thermodynamic uncertainty relations (TURs)~\cite{Barato2015, Gingrich2016, Pietzonka2016, Pietzonka2018}, which express a trade-off between the relative fluctuation of observables and entropy production. As for quantum information, it can provide tools and insights for furthering the understanding of nonclassical signatures in fluctuations in non-equilibrium systems governed by quantum dynamics, such as coherence and entanglement quantifiers~\cite{Baumgratz2014, Horodecki2009, Streltsov2017}.
In this sense, interdisciplinary approaches involving these two fields can enable promising research avenues, as indicated by recent literature on quantum fluctuation theorems~\cite{Esposito2009, Campisi2011, Batalhao2014, Kwon2019, DeChiara2022, Prech2023,  Zhang2024}, classical TUR violations~\cite{Ptaszynski2018, Agarwalla2018, Kalaee2021, PrechJohansson2023} and TURs including quantum corrections~\cite{Hasegawa2020, Vu2022}.

In the same vein, the work fluctuation-dissipation relation (FDR), a paradigmatic result in stochastic thermodynamics, was originally derived for classical systems close to equilibrium~\cite{Jarzysnki1997, Hendrix2001,Hermans1991}, but it has recently been generalized to encompass quantum dynamics~\cite{Miller2019,Scandi2020}.
Given a classical system coupled to a bath at temperature $T$ and subjected to a slow change of its Hamiltonian, the distribution for the work $W$ done in the process satisfies the work FDR, which is given by
\begin{equation}\label{WFDR}
   \langle W_{\text{diss}}\rangle - \frac{\beta}{2}{\rm Var}(W) = 0,
\end{equation}
where $W_{\text{diss}} \equiv W - \Delta F$, $\Delta F$ being the change in the equilibrium free energy, ${\rm Var}(W) \equiv \langle W^2 \rangle - \langle W \rangle^2$ is the variance of the work distribution, and $\beta = 1/k_{\text B} T$, $k_{\text B}$ being the Boltzmann constant. The work FDR implies that whenever dissipation occurs in classical systems close to equilibrium, fluctuations are produced.
Eq.~\eqref{WFDR} reflects the Gaussianity of the work distribution in slowly driven classical systems~\cite{Speck2004,Hoppenau2013}, and can be derived from the Jarzynski equality~\cite{Jarzysnki1997, Jarzynski2011}.
The work FDR has been experimentally tested and verified in a number of systems, including a colloidal particle in a feedback trap~\cite{Jun2014} and a semiconductor quantum dot system~\cite{Barker2022}.

The generalization of the work FDR protocol~\cite{Scandi2020} is built upon the discretization of a quasi-isothermal thermodynamic process into $N$ steps~\cite{Nulton1985, Crooks2007, Anders2013, Gallego2014} and the two-point measurement (TPM) scheme~\cite{Talkner2016}.
Given two fixed system Hamiltonians $H_I$ and $H_F$ as endpoints for the process, $N \gg 1$ ensures that it is sufficiently slow~\cite{Scandi2020}.
Before each step of the protocol, the system is in thermal equilibrium with the bath, and therefore in a Gibbs state. A first projective measurement in the system's energy eigenbasis is performed, and the outcome is recorded. Then, a quench on the Hamiltonian 
may generate coherence in the instantaneous system's energy eigenbasis. Finally, a second measurement in the new energy eigenbasis is performed and the result is recorded. The difference between the two recorded energy eigenvalues,
obtained via projective measurements, 
corresponds to the work in that step of the protocol, as defined by the TPM scheme.
Generation of coherence in the protocol leads to a correction of the work FDR in Eq.~\eqref{WFDR}~\cite{Scandi2020}.
This correction has recently been experimentally verified using a single trapped-ion qubit~\cite{Onishchenko2022}. These results have important consequences for the energetics of quantum information processing, e.g.~they imply that the generation of coherence during information erasure comes at the cost of increased dissipation~\cite{Miller2020,Vu2022b}.

Given that generation of coherence leads to a clear signature in the work statistics, it is logical to ask if other features of quantum dynamics can also give rise to characteristic thermodynamic signatures.
In this work, we address this question in the context of multipartite quantum systems, where quantum correlations play a role. Focussing on two-qubit systems, we show that entanglement generation during slow control protocols increases the mean dissipated work, via a universal correction to the classical FDR that is distinct from the contribution due to local coherences. 
Indeed, when only local (separable) unitary controls are applied, we show that this entanglement correction vanishes. In this case, we retrieve previous results, where the excess work variance arises from the generation of coherence in the local energy eigenbasis~\cite{Miller2019, Scandi2020, Onishchenko2022}.

\section{The work FDR protocol}

The work FDR protocol for quantum systems introduced in Ref.~\cite{Scandi2020} involves the discretization of a process with fixed endpoints, from the system Hamiltonian $H_I$ to $H_F$, into $N$ steps.
The system is in equilibrium with the bath at the beginning of the protocol.
In each step $i$, for $i = 0, \dots, N-1$, the Hamiltonian is $H_i$ at the beginning of the step, such that $H_0=H_I$ and $H_{N-1} = H_F$. 
The following two procedures are  performed in each step $i$:
\begin{enumerate}

\item  Quench on the Hamiltonian, changing it from $H_i$ to $H_{i+1}$. This process is very fast and does not change the state of the system. 

\item Thermalization procedure, in which the system's state is driven to a Gibbs state in the Hamiltonian's $H_{i+1}$ basis after interacting with the bath for a sufficiently long time,
\begin{equation}
    \pi_{i+1} = \frac{1}{Z_{i+1}} e^{-\beta H_{i+1}},
\end{equation}
where $Z_i = {\rm tr}\{ e^{-\beta H_{i}} \}$.

\end{enumerate}

As the system starts in equilibrium with the bath, its state at the beginning of the protocol is $\pi_0$.
In this way, the system is always in the Gibbs state $\pi_i$ before the $i$-th step.
A measurement of $\pi_{i}$ is then performed on the basis of $H_{i}$.  
A second measurement is performed on the basis of $H_{i+1}$ after the quench (i.e., after the completion of the step).
The difference between the two energy eigenvalues corresponds to the stochastic work in the step, $w_i$. Concretely, given discrete Hamiltonians $H_i = \sum_j E^{i}_{j}|E^{i}_{j}\rangle\langle E^{i}_{j}|$ and $H_{i+1} = \sum_j E^{i+1}_{j}|E^{i+1}_{j}\rangle\langle E^{i+1}_{j}|$, where $\{ \ket{E_j^i}\}$ and $\{ \ket{E_j^{i+1}}\}$ are eigenbases of $H_i$ and $H_{i+1}$ and $E^i_j$ and $E^{i+1}_j$ are eigenvalues, then the stochastic work will be $w_i = E_l^{i+1} - E_k^i$ when the outcomes $E_k^i$ and $E_l^{i+1}$ are obtained for the first and second measurement, respectively.

An experimental example of a quantum violation of the classical FDR was reported in Ref.~\cite{Onishchenko2022} using a trapped-ion qubit.
The initial and final Hamiltonians are $H_I = \epsilon\hat{\sigma}_z/2$ (we assume $\epsilon = 1$ throughout) and $H_F = \hat{U}^\dagger H_I \hat{U}$, with
\begin{equation}\label{Unitary}
\hat{U}=e^{-i\frac{\theta\hat{\sigma}_x}{2}}=\left(\begin{array}{cc}
\cos \frac{\theta}{2} & -\mathrm{i} \sin \frac{\theta}{2} \\
-\mathrm{i} \sin \frac{ \theta}{2} & \cos \frac{\theta}{2}
\end{array}\right),
\end{equation}
where $\hat{\sigma}_k$, $k = x, y, z$ are Pauli matrices. The protocol described above implies that, in each step $i$, a rotation of $\Delta \theta  = \theta/N$ is performed, such that $\hat{U}_i = e^{-i\Delta \theta \hat{\sigma}_x}$ and
\begin{equation}\label{DUQS}
    H_i\longmapsto\hat{U}^{\dagger}_iH_i\hat{U}_i \equiv H_{i+1}.
\end{equation}
Note that, in this case, the eigenvalues do not change as
\begin{equation}
\begin{split}
H_{i+1} \equiv \sum_j E^{i+1}_{j}|E^{i+1}_{j}\rangle\langle E^{i+1}_{j}| \\
= \sum_j E^{i}_{j}\hat{U}^{\dagger}_i|E^{i}_{j}\rangle\langle E^{i}_{j}|\hat{U}_i &= \hat{U}^{\dagger}_iH_i\hat{U}_i.
\end{split}
\end{equation}
The probability of getting the work $w_i = E_l^{i+1} - E_k^i$ is given by
\begin{equation}\label{prob}
    P_i(w_i) = \sum_{E^{i+1}_{l}-E^{i}_{k} = w_i} \langle E^{i}_k|\pi_i|E^{i}_k\rangle |\langle E^{i}_{k}|E^{i+1}_{l}\rangle|^2.
\end{equation}
In the computational basis, we have that
\begin{equation}
    \pi_i=\frac{1}{Z}\left(\begin{array}{cc}
1 &0 \\
0 & e^{-\beta}
\end{array}\right),
\end{equation}
where $Z = (1+e^{-\beta})^{-1}$.
The outcomes for the energy measurements in the instantaneous Hamiltonian basis, before and after the quench, are defined as either $0$ or $1$, when the system is in the ground or excited state, respectively. In this way, the possible values for the work in the step $i$ are $w_i = -1$, $w_i = 0$ and $w_i = 1$.
Thus, we can write
\begin{align}
    \langle w_i \rangle = \sum_{w_i} w_i P_i(w_i), \\
    \langle w_i^2 \rangle = \sum_{w_i} w_i^2 P_i(w_i).
\end{align}
After all $N$ steps, the total work is given by the sum of the work obtained in all steps, $W = \sum_i^N w_i$, which is also a stochastic quantity.
Note that the thermalization resets the system state at the beginning of the step, making the protocol effectively Markovian~\cite{Onishchenko2022}. Therefore, all $P_i(w_i)$ in each of the $N$ steps are statistically identical and independent, which allows us to write 
\begin{align}
    \langle W \rangle = N\langle w_i \rangle, \label{Wcoh}  \\
    \langle W^2 \rangle = N\langle w_i^2 \rangle. \label{Wcoh2}
\end{align}
Since the initial and final Hamiltonians are connected via a unitary transformation~\eqref{Unitary}, the change in free energy is zero, i.e., $\Delta F = 0$.

The correction to the work FDR is defined as
\begin{equation}\label{Q}
    Q \equiv  \langle W_{\text{diss}}\rangle - \frac{\beta}{2}{\rm Var(W)}.
\end{equation}
In the case above, using Eqs.~\eqref{Wcoh} and~\eqref{Wcoh2}, the correction is found to be~\cite{Onishchenko2022}
\begin{multline}
    Q = N \sin^2\left(\frac{\Delta \theta}{2}\right)\Bigg[ \frac{\beta}{2}\left(1-\sin^2\left(\frac{\Delta \theta}{2}\right)\tanh^2 (\beta/2)\right)  \\ - \tanh (\beta/2)\Bigg].
\end{multline}
As the process is slow and close to equilibrium, $N$ must be large. This implies that $\Delta \theta \ll 1$. By expanding $Q$ for small $\Delta \theta$, the correction can be approximated as 
\begin{align}\label{singleQ}
Q \approx 
N \frac{(\Delta\theta)^2}{4}f(\beta), 
\end{align}
where 
\begin{equation}\label{f}
    f(\beta) \equiv \frac{\beta}{2} - \tanh (\beta / 2).
\end{equation}
We note that $f(\beta)$ is a monotonic function, which goes to zero when $\beta$ goes to zero. This implies that the correction of the work FDR due to the generation of coherence tends to vanish as the temperature of the bath is increased.


\section{Multipartite work FDR protocol}

We now generalize the work FDR protocol to multipartite systems.
To keep the presentation simple, we focus on a bipartite system. However, the protocol for multipartite system follows directly by considering a larger number of systems.

%
%
Consider two systems, namely $S_A$ and $S_B$, coupled to a bath at inverse temperature $\beta$. 
We focus on a single step $i$ of the protocol, as introduced in the previous section, and therefore we omit the index $i$ throughout, as all $N$ steps are identical and statistically independent.
At the beginning of the step, the Hamiltonian of each subsystem is $H_A$ and $H_B$.
Given that $\{\ket{E_j^A}\}$ and $\{\ket{E_k^B}\}$ are the eigenbasis of $H_A$ and $H_B$, then $\{\ket{E_j^A} \ket{E_k^B}\}$ is the eigenbasis of the total Hamiltonian $H_{AB} = H_A \otimes H_B$, and the corresponding projectors associated with the first energy measurement are $\hat{P}_{jk} =  |E_j^A\rangle  |E^B_k\rangle\langle E_j^A| \langle E_k^B|$. 
After local quenches on $H_A$ and $H_B$, the new Hamiltonian will be $H_{AB}^\prime = H_A^\prime \otimes H_B^\prime$, and the projectors associated with the new measurement of the TPM scheme are given by $\hat{P}_{jk}^\prime = \hat{P}^\prime_j \otimes \hat{P}^\prime_k$, where $\hat{P}_j^\prime$ and $\hat{P}_k^\prime$ are local projectors on the local Hamiltonians' eigenbases.

To include the possibility of interaction between the two systems, and therefore entanglement generation,
we add an extra procedure within each step (procedure 2 below), corresponding to a unitary transformation of the system's state,
the other two procedures being essentially the same as before (procedures 1 and 3):
\begin{enumerate}

     \item Quench on the Hamiltonian $H_{AB} = H_A\otimes H_B$, which become $H_{AB}^\prime = H_A^\prime \otimes H_B^\prime$.

    \item Unitary evolution of the global system's state $\rho_{AB}$, 
    \begin{equation}
       \rho_{AB} \longmapsto \hat{R} \rho_{AB} \hat{R}^\dagger,
    \end{equation}
    where $\hat{R}$ is a unitary operator. Entanglement may be generated in this process for interacting systems.
   
    \item The system is thermalized to a Gibbs state in the basis of $H_{AB} = H_A \otimes H_B$, at inverse temperature $\beta$: $\pi_{AB} = \pi_A \otimes \pi_B$, with
    \begin{equation}
        \pi_{A} = \frac{e^{-\beta H_A}}{Z_A}, \ \ \pi_B = \frac{e^{-\beta H_B}}{Z_B},
    \end{equation}
    where $Z_A = {\rm tr}\{e^{-\beta H_A}\}$ and $Z_B = {\rm tr}\{e^{-\beta H_B}\}$.
\end{enumerate}

\section{Two-qubit example}
\begin{figure}[]
\centering
\includegraphics[clip,width=0.95\columnwidth]{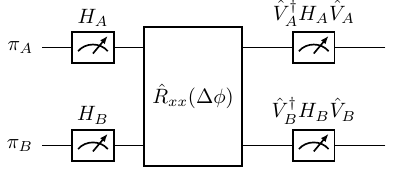}
\caption{Sketch of the TPM scheme in one step of the work FDR protocol for a two-qubit system.
The system state $\pi_A \otimes \pi_B$ is projectively measured on the basis of $H_{AB} = H_A 
\otimes H_B$.
The quench is then applied to the Hamiltonian, which becomes $H_{AB}^\prime = H_A^\prime \otimes H_B^\prime$, while a unitary transformation described by $\hat{R} = \hat{R}_{xx}$ acts on the system state. Finally, the second measurement in the TPM scheme is performed on the basis of $H_{AB}^\prime$.}
\label{doublequbit}
\end{figure}

To illustrate how entanglement generation can affect work fluctuations in slow processes, we examine an example with two qubit systems using our new protocol.
We consider $H_A = H_B = \sigma_z/2$, the eigenstates of $\sigma_z$ being $\ket{0}$ and $\ket{1}$.
As in the example for one qubit system discussed above, we consider local quenches given by
\begin{equation}\label{LocalU}
\hat{V}_A = \hat{V}_B = e^{-\mathrm{i}\frac{\Delta \theta}{2}\hat{\sigma}_x}.
\end{equation}
The local Hamiltonians are therefore transformed as follows,
\begin{align}
    H_{A}\longmapsto\hat{V}_A^{\dagger}H_{A}\hat{V}_A = H^{\prime}_A, \\
    H_{B}\longmapsto \hat{V}_B^{\dagger}H_{B}\hat{V}_B = H^{\prime}_B.
\end{align}
The global unitary $\hat{R}$ acting on the system's state will be considered as the following global rotation,
\begin{equation}\label{RXX}
\hat{R}=\hat{R}_{xx}(\Delta 
\phi) \equiv e^{-\mathrm{i}\frac{\Delta \phi}{2} \hat{\sigma}_x\otimes \hat{\sigma}_x }.
\end{equation}
Note that we introduced a new parameter, $\Delta \phi = \phi/N$, which discretizes the total rotation angle $\phi$, in the same way that $\Delta \theta = \theta / N$ discretizes $\theta$. 
In this case, procedures 2 and 3 together can be seen as single discrete unitary transformation, as defined in Ref.~\cite{Anders2013}.
Before the step, the system's state is
$\pi_{AB} = \pi_A \otimes \pi_B$. Since $H_A = H_B = \hat{\sigma}_z/2$, we have that
\begin{equation}
    \pi_A = \pi_B = \frac{1}{Z}\left(\begin{array}{cc}
1 &0 \\
0 & e^{-\beta}
\end{array}\right).
\end{equation}
The eigenstates of the Hamiltonian $H_{AB}$ are $\ket{E_0} \equiv \ket{00}$, $\ket{E_1} \equiv \ket{10}$, $\ket{E_2} \equiv \ket{01}$ and $\ket{E_3} \equiv \ket{11}$ with correspondent eigenvalues $E_0 = 0$, $E_1 = E_2 = 1$, and $E_3 = 2$, and the eigenstates of $H_{AB}^\prime$ will be denoted by $\ket{E_0^\prime} \equiv \hat{V}^\dagger\ket{00}$, $\ket{E_1^\prime} \equiv \hat{V}^\dagger\ket{10}$, $\ket{E_2^\prime} 
\equiv\hat{V}^\dagger\ket{01}$ and $\ket{E_3^\prime}\equiv\hat{V}^\dagger\ket{11}$, with eigenvalues $E_0^\prime = 0$, $E_1^\prime = E_2^\prime = 1$, and $E_3^\prime = 2$.  We sketch a step of the protocol in Fig.~\ref{doublequbit}, including the measurements in the TPM scheme on the respective local Hamiltonians eigenbasis, before and after the quench. 

The probability of getting the work $w$ in a step of the protocol can therefore be expressed as
\begin{equation}\label{entanglementwork}
P(w) = \sum_{E^{\prime}_{l}-E_{u} = w} \langle E_u|\pi_{AB}|E_u\rangle |\langle E_{l}^\prime|\hat{R}|E_{u}\rangle|^2.
\end{equation}
There are now four possible values for the work $w = E_l^\prime - E_u$ in each step: $w = -2$, $w=1$, $w=0$, $w=1$ and $w=2$.
As before, given the total work $W$ after all $N$ steps, we have that
\begin{align}
    \langle W\rangle = N \langle w \rangle, \label{Wcumm1} \\
    \langle W^2\rangle = N \langle w^2 \rangle. \label{Wcumm2}
\end{align}
Since $N$ must be large for a process in the slow driving regime, using Eq.~\eqref{entanglementwork} and the small angle approximation, we obtain from Eq.~\eqref{Q} the following correction,
\begin{equation}\label{correction_entanglement}
Q_{AB} \approx N\left[\frac{(\Delta \theta)^2}{2}f(\beta)
    + \frac{(\Delta \phi)^2}{2}g(\beta)\right],
\end{equation}
where $f(\beta)$ is given in Eq.~\eqref{f} and
\begin{equation}\label{g}
    g(\beta) \equiv \frac{\beta}{1 + {\rm sech}\,\beta} - \tanh{(\beta/2)}.
\end{equation}
We note that both functions are monotonic and vanish for large temperature $T$  (or small $\beta$). 
Furthermore, in the absence of $\hat{R}$ (for $\Delta \phi = 0$), Eq.~\eqref{correction_entanglement} becomes
\begin{equation}\label{noentangl}
    Q_{AB} = 2Q = N\left[\frac{(\Delta \theta)^2}{2}f(\beta)\right],
\end{equation}
where $Q$ is the correction for a single qubit given in Eq.~\eqref{singleQ}. 
This is consistent with the fact that, in this case, only local coherences contribute to $Q_{AB}$, the factor of $2$ accounting for the fact that the system is now constituted of two qubits.
The function $f(\beta)$, therefore, expresses the dependence of the correction on the bath temperature due to coherence generation in the local instantaneous energy eigenbasis.

 To see if entanglement is generated in the protocol, we employ an entanglement quantifier. Given a a state $\rho_{AB}$, an entanglement monotone is given by its {\it negativity}~\cite{Vidal2002},
\begin{equation}
    \mathcal{N}(\rho_{AB}) \equiv \frac{||\rho_{AB}^{\Gamma_A}||_{1} - 1}{2} \geq 0, 
\end{equation}
where $\rho_{AB}^{\Gamma_A}$ is the partial transpose of $\rho_{AB}$ with respect to subsystem A, and $||X||_1 = {\rm tr}\sqrt{X^\dagger X}$. To use the negativity as a quantifier of the entanglement generated in the protocol, we consider each of the 4 possible eigenstates $ \ket{E_u}$, with $u = 0,1,2,3$, which result from the first energy measurement, as inputs to which
the unitary operation $\hat{R}$ will be applied. We calculate the negativity of each possible resulting entangled state, $\hat{R}\ket{E_u}\bra{E_u}\hat{R}^\dagger$ (see Appendix~\ref{AppendixC}).
Since $\Delta \phi$ is very small, we expand the negativity in terms of this parameter, which to first order in $\Delta \phi$ is given by
\begin{equation}
    \mathcal{N}(\hat{R} \ket{E_u}\bra{E_u}\hat{R}^\dagger) \approx \Delta \phi,
\end{equation}
for all $u$. This implies that entanglement is generated in each step of the protocol, vanishing only for $\Delta \phi = 0$. We can therefore conclude that the second term in the right hand side of Eq.~\eqref{correction_entanglement} will be different from zero as long as entanglement is generated in the protocol. The function, $g(\beta)$, in turn, captures the temperature dependence of the entanglement signature in the work fluctuations.

The difference between $g(\beta)$ and $f(\beta)$ in Eq.~\eqref{correction_entanglement} is given by
\begin{equation}
    g(\beta) - f(\beta) = \frac{\beta}{2} \tanh^2(\beta/2) \geq 0.
\end{equation}
For $\beta \ll 1$, we find that
\begin{equation}
    g(\beta) - f(\beta) = \mathcal{O}(\beta^3).
\end{equation}
Therefore, for small $\beta$ or large bath temperature, the two functions become indistinguishable and it becomes harder to see a distinct entanglement signature in the dissipated work. By contrast, in the low-temperature limit we have $\lim_{\beta\to\infty} g(\beta)/f(\beta) = 2$, i.e.~the quantum correction from entanglement is twice as large as the one from local coherences.

 
\section{Work FDR with generalized two-qubit entangler}

By choosing a specific entangling unitary, given by $\hat{R}_{xx}$ in Eq.~\eqref{RXX}, we showed that entanglement generation leads to a separate correction to the work FDR. 
Here, we generalize this result by considering a general two-qubit unitary operator in SU(4)~\cite{Rezhakani2004}
\begin{equation}\label{entangler}
    \hat{O}=[\hat{A}_1 \otimes \hat{B}_1] e^{-\mathrm{i}\left(c_1 \hat{\sigma}_x \otimes \hat{\sigma}_x+c_2 \hat{\sigma}_y \otimes \hat{\sigma}_y+c_3 \hat{\sigma}_z \otimes \hat{\sigma}_z\right)}[\hat{A}_2 \otimes \hat{B}_2],
\end{equation}
where $A_i$, $B_i$ $\in$ SU(2) are arbitrary single qubit unitary operations. 
As $A_i$ and $B_i$ cannot generate entanglement, we define the unitary operator $\hat{R}$ as
\begin{equation}\label{entangler1}
    \hat{R} \equiv e^{-\mathrm{i}\left(c_1 \hat{\sigma}_x \otimes \hat{\sigma}_x+c_2 \hat{\sigma}_y \otimes \hat{\sigma}_y+c_3 \hat{\sigma}_z \otimes \hat{\sigma}_z\right)}.
\end{equation}
We assume that all parameters $c_1, c_2, c_3$ scale with $N^{-1}$, just as $\Delta \theta$ does. This assumption follows the same logic as presented before for Eq.~\eqref{RXX}, where $\Delta \phi \propto N^{-1}$, as the entire protocol is discretized into $N$ statistically independent steps. 
Using Eq.~\eqref{entanglementwork}, we follow similar steps as described in the previous section (see Appendix~\ref{AppendixA} for details) to find the following correction to the work FDR in the slow driving regime,
\begin{equation}\label{QABc1c2}
    Q_{AB} \approx N\left[\frac{(\Delta \theta)^2}{2}f(\beta)
    + 2(c_1 - c_2)^2 g(\beta)\right].
\end{equation}
Note that Eq.~\eqref{RXX} is a particular case of the general unitary operator in Eq.~\eqref{entangler1}, for which $c_1 = \Delta \phi /2 = \phi/2N$ and $c_2 = c_3 = 0$. Thus, the correction in Eq.~\eqref{correction_entanglement} is obtained from Eq.~\eqref{QABc1c2} for these particular values of $c_1$, $c_2$ and $c_3$.

As in the example of the previous section, we calculate the negativities $\mathcal{N}_u \equiv \mathcal{N}(\hat{R} \ket{E_u}\bra{E_u}\hat{R}^\dagger)$, $\hat{R}$ being given in Eq.~\eqref{entangler1} (see Appendix~\ref{AppendixC}).
For $N$ large, we get 
\begin{align}
    \mathcal{N}_1 = \mathcal{N}_2 \approx c_1 + c_2, \label{negativitydeg} \\
    \mathcal{N}_0 = \mathcal{N}_3 \approx c_1 - c_2. \label{negativitynondeg}
\end{align}
Therefore, entanglement is always generated in the protocol even if $c_1 = c_2$, but in this case the contribution proportional to $g(\beta)$ in Eq.~\eqref{QABc1c2} vanishes. This is due to the fact that the work probability distribution depends only on the difference between the parameters $c_1-c_2$, as shown in the SI.
In this way, entanglement generated within the degenerate subspace of local energy eigenstates, $\{\ket{01},\ket{10}\}$, does not contribute to the correction $g(\beta)$. Only entanglement involving the non-degenerate eigenstates leads to a signature in the work statistics.

Finally, we consider arbitrary separable unitaries $\hat{R}$, 
\begin{equation}\label{Rseparable}
    \hat{R} = \hat{R}_A \otimes \hat{R}_B,
\end{equation}
where $\hat{R}_A$ and $\hat{R}_B$ are local arbitrary unitary operators.
In Appendix~\ref{AppendixB}, we show that for arbitrary $\hat{R}_A$ and $\hat{R}_B$, the contribution proportional to $g(\beta)$ in Eq.~\eqref{QABc1c2}  disappears. The only contribution to the correction $Q_{AB}$, in this case, is due to generation of coherence in the local Hamiltonians basis, which is proportional to $f(\beta)$.

\section{Conclusion}

We proposed a generalization of the work FDR protocol for multipartite systems, going beyond previous work~\cite{Scandi2020, Onishchenko2022} by permitting the possibility of quantum correlations to develop between locally thermalizing subsystems.
In particular, our framework allows for entanglement generation during the driving stage, while assuming that the environment brings the subsystems to local thermal equilibrium after each small time step. Such local equilibration is a good description of thermal dissipation for weakly coupled subsystems~\cite{Rivas2010,Gonzalez2017,Hofer2017}, e.g.~see Ref.~\cite{Potts2021} for an explicit construction of a local ``thermodynamic'' Hamiltonian. 
 In the slow-driving regime, we derived
 the correction to the work FDR considering general entangling operations acting on two-qubit systems. 
 Using the negativity as an entanglement quantifier, we identified an additive correction to the dissipated work due to entanglement generation, which dominates the quantum correction due to local coherences at low temperatures.

The fact that quantum entanglement increases dissipation provides further evidence that quantum effects are detrimental to the efficiency of thermodynamic tasks in the near-equilibrium regime~\cite{Brandner2017,Miller2019,Scandi2020,Miller2020,Miller2021,Vu2022b}. Notably, however, we found that only entanglement generation that involves non-degenerate local energy eigenstates plays a detrimental role. This contrasts with other recent findings that energetic (i.e.~non-degenerate) coherence can be a thermodynamic resource outside of the slow-driving regime~\cite{Hammam2021,Almanza-Marrero2024}. It is also interesting to compare our results with those of Rolandi et al.~\cite{Rolandi2023}, who recently showed that collective operations may greatly reduce dissipation in slow driving protocols. In that case, however, the minimally dissipative protocols involve Hamiltonians with classical (mutually commuting) interactions, which cannot generate entanglement. This is fully consistent with our conclusion that entanglement always increases dissipation in the slow-driving regime.

Our findings imply an increased thermodynamic cost for quantum information processing protocols where entanglement is unavoidably generated, e.g.~cross-talk between superconducting qubits, which is an important source of noise in near-term quantum devices~\cite{Arute2019,vandenBerg2023,Kim2023a}. We also expect that the relative simplicity of our protocol will make it amenable to implementation in a controlled setting, thus unlocking further experimental exploration of quantum signatures in the nonequilibrium thermodynamics of small systems.

\section*{Acknowledgements}
We thank P. Samuelsson and S. Brattegard for insightful discussions, and thank H. Miller and M. Perarnau-Llobet for their comments on the manuscript. M.T.M. is supported by a Royal Society-SFI University Research Fellowship (Grant No.~URF\textbackslash R1\textbackslash 221571).
S.V.M. acknowledges support from the Knut and Alice Wallenberg Foundation (Project No.~2016-0089).
This project is co-funded by the European Union (Quantum Flagship project ASPECTS, Grant Agreement No.~101080167). Views and opinions expressed are however those of the authors only and do not necessarily reflect those of the European Union, Research Executive Agency or UKRI. Neither the European Union nor UKRI can be held responsible for them.

\appendix

\section{Work FDR correction for general two-qubit entangling unitaries}\label{AppendixA}

Here, we derive the work FDR correction in the bipartite work FDR protocol with general entangling unitaries given in Eq.~\eqref{entangler1}.
In the TPM scheme for systems A and B, we define $P_{ij,mn}$ as the probability distribution for outcomes $i$ and $j$ from the measurements of $H_A$ and $H_B$, while $m$ and $n$ are the outcomes of the measurements of $H_A^\prime$ and $H_B^\prime$ (after the quench). As the eigenvalues do not change due to the unitary quench, we have that $i$, $j$, $m$ and $n$ can only take the values $0$ or $1$.
The work in one step of the protocol can, therefore, be written as $ w = (m + n) - (i + j)$. 
For instance, $P_{00,01}$ represents the probability of obtaining $i,j = 0$ at the first measurement, followed by obtaining $m =0$ and $n=1$ at the second measurement. The work in this scenario, as determined by the TPM scheme, is $w = 1$.
Considering all combinations of outcomes such that $w=1$, the corresponding probability of getting this value for work will be given by
\begin{equation}
    P(w=1) = P_{00,01} + P_{00,10} + P_{01,11} + P_{10,11}.
\end{equation}
Generally, from Eq. (\ref{entanglementwork}), we have that
\begin{equation}
    P(w) = \sum_{E^{\prime}_{l}-E_{k} = w} \langle E_k|\pi_i|E_k\rangle |\langle E_{l}|(\hat{V}_A\otimes \hat{V}_B)\hat{R}|E_{k}\rangle|^2.
\end{equation}

We recall that, in the computational basis of both qubits, $\{|E_k\rangle \}= \{|00\rangle, |01\rangle,|10\rangle,|11\rangle \}.$
Given that ${\rm tr}\{A|\psi\rangle\langle\phi|\}=\langle\phi | A |\psi\rangle$, we can write
\begin{widetext}
\begin{align}\label{proj}
\langle E_k|\pi_i|E_k\rangle |\langle E_{l}|(\hat{V}_A\otimes \hat{V}_B)\hat{R}|E_{k}\rangle|^2= \langle ij|\pi_i|ij\rangle |\langle mn|(\hat{V}_A\otimes \hat{V}_B)\hat{R}|ij\rangle|^2 =
 {\rm tr}\{M^{ij}\pi_i\}{\rm tr}\{M^{mn}\rho^{ij}\} = P_{ij,mn}.
\end{align}
\end{widetext}
In the expression above, 
$M^{ij} = |ij\rangle\langle ij| $ and 
$M^{mn} = |mn\rangle\langle mn| $ are projectors on the computational basis of both qubits, $\rho^{ij} =(\hat{V}_A^\dagger\otimes \hat{V}_B^\dagger)\hat{R}|ij\rangle \langle ij|\hat{R}^\dagger(\hat{V}_A\otimes \hat{V}_B)$ and $\pi_i$ is the thermal state in the computational basis of both qubits,
\begin{equation}
\pi_i =\frac{1}{Z} \left(\begin{array}{cccc}
1 & 0 & 0 & 0 \\
0 & e^{-\beta} & 0 & 0 \\
0 & 0 & e^{-\beta} & 0 \\
0 & 0 & 0 & e^{-2\beta}
\end{array}\right).
\end{equation} 
In this way, we can write Eq. (\ref{entanglementwork}) as
\begin{widetext}
\begin{equation}
    P(w) = \sum_{(m+n)-(i+j) = w} \langle ij|\pi_i|ij\rangle |\langle mn|(\hat{V}_A\otimes \hat{V}_B)\hat{R}|ij\rangle|^2 
    = \sum_{(m+n)-(i+j) = w} {\rm tr}\{M^{ij}\pi_i\}{\rm tr}\{M^{mn}\rho^{ij}\}.
\end{equation}

Given $\hat{V}$ and $\hat{R}$, we can now work out all the elements $P_{ij,mn}$ needed to characterize the entire distribution $P(w)$. As an example, to calculate $P_{00,01}$, we use

\begin{equation}\label{energyeigen}
    M^{00} = \left(\begin{array}{cccc}
1 & 0 & 0 & 0 \\
0 & 0 & 0 & 0 \\
0 & 0 &  0& 0 \\
0 & 0 & 0 & 0
\end{array}\right),
\quad
M^{01} = \left(\begin{array}{cccc}
0 & 0 & 0 & 0 \\
0 & 1 & 0 & 0 \\
0 & 0 & 0 & 0 \\
0 & 0 & 0 & 0
\end{array}\right),
\end{equation}
and
\begin{equation}
    \rho^{00} = (\hat{V}_A\otimes \hat{V}_B)\hat{R}|00\rangle \langle 00|\hat{R}^\dagger(\hat{V}_A^\dagger\otimes \hat{V}_B^\dagger).
\end{equation}
The diagonal terms of $\rho^{00}$ are given by
\begin{align}
\rho^{00}_{11} &= \frac{1}{8}(3+ 2\cos(2c_1 -2c_2-\Delta\theta) + \cos(2\Delta\theta) + 2\cos(2c_1 -2c_2 + \Delta\theta)), \\
\rho^{00}_{22} &= \rho^{00}_{33} = \frac{\sin^2(\Delta\theta)}{4}, \\
\rho^{00}_{44} &= \cos^4\left(\frac{\Delta\theta}{2}\right) \sin^2(c_1 - c_2)  - \cos^2(c_1- c_2) \sin^4\left(\frac{\Delta\theta}{2}\right).
\end{align}

We can check that ${\rm tr}\{\rho^{00}\} = 1$. Finally, we can find that
\begin{equation}
  P_{00,01}   = {\rm tr}\{M^{00}\pi_i\}{\rm tr}\{M^{01}\rho^{00}\} = \frac{e^{2\beta}\sin(\Delta \theta)^2}{4(1+e^{\beta})^2}.
\end{equation}
From this and the other $P_{ij,mn}$ terms, we are able to evaluate the work distribution $P(w)$:
\begin{align}
P(-2) &= \frac{\cos^4\left(\frac{\Delta\theta}{2}\right) \sin^2(c_1-c_2) + \sin^4\left(\frac{\Delta\theta}{2}\right) \cos^2(c_1-c_2)}{\left(e^{\beta}+1\right)^2}, \\
P(-1) &= \frac{\sin^2(\Delta\theta)}{2 \left(e^{\beta}+1\right)}, \label{P-1}\\
P(1)  &= \frac{e^{\beta} \sin^2(\Delta\theta)}{2 \left(e^{\beta}+1\right)}, \label{P1}\\
P(2)  &= \frac{e^{2{\beta}} \left(\cos^4\left(\frac{\Delta\theta}{2}\right) \sin^2(c_1-c_2)+\sin^4\left(\frac{\Delta\theta}{2}\right) \cos^2(c_1-c_2)\right)}{\left(e^{\beta}+1\right)^2}.
\end{align}
\end{widetext}
Note that we do not write $P(0)$ out since it will not contribute when calculating the work cumulants we need.
As discussed before, the free energy change is zero, $\Delta F = 0$, since the quench is a unitary transformation. Thus, all we need are the cumulants of work to find the correction function for the entire protocol,
\begin{equation}\label{QABsupp}
    Q_{AB} =  \langle W_{\text{diss}}\rangle - \frac{\beta}{2}{\rm Var(W)}.
\end{equation}
Using Eqs.~\eqref{Wcumm1} and~\eqref{Wcumm2} in the main text, we get
\begin{widetext}
\begin{align}
\frac{Q_{AB}}{N} = \frac{1}{2} \beta &\bigg \{-\bigg [\frac{2 e^{2 \beta} \left(\cos^4\left(\frac{\Delta\theta}{2}\right) \sin^2(c_1-c_2) + \sin^4\left(\frac{\Delta\theta}{2}\right) \cos^2(c_1-c_2)\right)}{\left(e^\beta+1\right)^2} \notag \\
&-\frac{2 \left(\cos^4\left(\frac{\Delta\theta}{2}\right) \sin^2(c_1-c_2) + \sin^4\left(\frac{\Delta\theta}{2}\right) \cos^2(c_1-c_2)\right)}{\left(e^\beta+1\right)^2}+\frac{(e^{2 \beta}-1) \sin^2(\Delta\theta)}{2 \left(e^\beta+1\right)^2}\bigg]^2 \notag \\
&+\frac{4 e^{2 \beta} \left(\cos^4\left(\frac{\Delta\theta}{2}\right) \sin^2(c_1-c_2) + \sin^4\left(\frac{\Delta\theta}{2}\right) \cos^2(c_1-c_2)\right)}{\left(e^\beta+1\right)^2}+\frac{4 \left(\cos^4\left(\frac{\Delta\theta}{2}\right) \sin^2(c_1-c_2) + \sin^4\left(\frac{\Delta\theta}{2}\right) \cos^2(c_1-c_2)\right)}{\left(e^\beta+1\right)^2} \notag \\
&+\frac{\sin^2(\Delta\theta)}{2}\bigg\} +\frac{2 \left(\cos^4\left(\frac{\Delta\theta}{2}\right) \sin^2(c_1-c_2) + \sin^4\left(\frac{\Delta\theta}{2}\right) \cos^2(c_1-c_2)\right)}{\left(e^\beta+1\right)^2} \notag \\
&-\frac{2 e^{2 \beta} \left(\cos^4\left(\frac{\Delta\theta}{2}\right) \sin^2(c_1-c_2) + \sin^4\left(\frac{\Delta\theta}{2}\right) \cos^2(c_1-c_2)\right)}{\left(e^\beta+1\right)^2} -\frac{(e^{2 \beta}-1) \sin^2(\Delta\theta)}{2 \left(e^\beta+1\right)^2}
\end{align}
\end{widetext}
As discussed in the main text, $c_1$, $c_2$, $c_3$ and $\Delta \theta$ are all proportional to $N^{-1}$, where $N$ is the number of the steps in the entire protocol. As the slow driving regime corresponds to large $N$, $c_1 - c_2$ and $\Delta \theta$ will be small, so that we can expand the expression above to leading order,
\begin{equation}
    Q_{AB} \approx N\left[\frac{(\Delta \theta)^2}{2}f(\beta)
    + 2(c_1 - c_2)^2 g(\beta)\right],
\end{equation}
where $f(\beta)$ and $g(\beta)$ are given in Eqs.~\eqref{f} and~\eqref{g} in the main text. 

\section{Work FDR correction for arbitrary separable unitaries}\label{AppendixB}
In Eq.~\eqref{entangler} in the main text, we considered $\hat{R}$ to be an inseparable unitary operator. In this case, we found that generation of entanglement can lead to an extra term in the correction of the work FDR.
Here, we set $\hat{R}$ to be an arbitrary separable unitary operator. We will show that separable operators cannot lead to that extra term in the correction, which supplements our conclusion that it corresponds to a signature of entanglement generation in the protocol.

We start by considering $\hat{R}$ to be an arbitrary separable two-qubit unitary given by
\begin{equation}
    \hat{R} = \hat{R}_{\text S} \equiv \hat{R}_A \otimes \hat{R}_B,
\end{equation}
where $\hat{R}_A$ and $\hat{R}_B$ are arbitrary local unitary operators for systems $A$ and $B$.
Note that $\hat{R}_A$ and $\hat{R}_B$ can be decomposed in terms of the following operators,
\begin{widetext}
\begin{align}
    R_X(\theta) &= e^{-i\frac{\theta}{2}\sigma_x} = \left(\begin{array}{cc}
\cos \frac{\theta}{2} & -i \sin \frac{\theta}{2} \\
-i \sin \frac{\theta}{2} & \cos \frac{\theta}{2}
\end{array}\right), \\
    R_Y(\theta) &= e^{-i\frac{\theta}{2}\sigma_y}= 
    \left(\begin{array}{cc}
\cos \frac{\theta}{2} & -\sin \frac{\theta}{2} \\
\sin \frac{\theta}{2} & \cos \frac{\theta}{2}
\end{array}\right),\\
  R_Z(\theta) &= e^{-i\frac{\theta}{2}\sigma_z}= 
  \left(\begin{array}{cc}
e^{-i \frac{\theta}{2}} & 0 \\
0 & e^{i \frac{\theta}{2}}
\end{array}\right).
\end{align}
We now decompose $\hat{R}_{\text S}$ into the X-Z-X form.
We note that the last $\hat{R}_X$ in the decomposition for each qubit can be absorbed into the quench $\hat{V}_A$ and $\hat{V}_B$ as $\hat{V}_A = \hat{V}_B = e^{-i\Delta\theta\sigma_x}$. Thus, we can write
\begin{equation}
    \hat{R}_{\text S}(c,l,m,n) = \hat{R}_X(c)\hat{R}_Z(l) \otimes \hat{R}_X(m)\hat{R}_Z(n),
\end{equation}
so that
\begin{equation}
    P(w) = \sum_{E^{\prime}_{l}-E^{i}_{k} = \omega} \langle E_k|\pi_i|E_k\rangle |\langle E_{l}|(\hat{V}_A\otimes \hat{V}_B)\hat{R}_{\text S}|E_{k}\rangle|^2.
\end{equation}

Proceeding as in Appendix~\ref{AppendixA}, we can then evaluate all the terms of the probability distribution
 $P(w)$, which are given by
\begin{align}
P(-2) &= \frac{\sin ^2\left(\frac{c+\Delta\theta}{2}\right) \sin ^2\left(\frac{m+\Delta\theta}{2}\right)}{\left(e^{\beta}+1\right)^2}, \\
P(-1) &= \frac{\sin ^2\left(\frac{c+\Delta\theta}{2}\right) \cos ^2\left(\frac{m+\Delta\theta}{2}\right)}{\left(e^\beta+1\right)^2}+\frac{e^\beta \cos ^2\left(\frac{c+\Delta\theta}{2}\right) \sin ^2\left(\frac{m+\Delta\theta}{2}\right)}{\left(e^\beta+1\right)^2}+\frac{\cos ^2\left(\frac{c+\Delta\theta}{2}\right) \sin ^2\left(\frac{m+\Delta\theta}{2}\right)}{\left(e^\beta+1\right)^2}\notag\\
&+\frac{e^\beta \sin \left(\frac{c+\Delta\theta}{2}\right) \cos \left(\frac{m+\Delta\theta}{2}\right) \left(-\frac{1}{2} \sin \left(\frac{c}{2}\right) \sin \left(\frac{m}{2}\right) \sin (\Delta\theta)+\sin \left(\frac{c}{2}\right) \cos \left(\frac{m}{2}\right) \cos ^2\left(\frac{\Delta\theta}{2}\right)\right)}{\left(e^\beta+1\right)^2}\notag\\
&+\frac{e^\beta \sin \left(\frac{c+\Delta\theta}{2}\right) \cos \left(\frac{m+\Delta\theta}{2}\right)\cos \left(\frac{c}{2}\right) \sin \left(\frac{\Delta\theta}{2}\right) \cos \left(\frac{m+\Delta\theta}{2}\right)}{\left(e^\beta+1\right)^2},\\
P(1)  &= \frac{e^\beta \sin ^2\left(\frac{c+\Delta\theta}{2}\right) \cos ^2\left(\frac{m+\Delta\theta}{2}\right)}{\left(e^\beta+1\right)^2}+\frac{e^{2 \beta} \sin ^2\left(\frac{c+\Delta\theta}{2}\right) \cos ^2\left(\frac{m+\Delta\theta}{2}\right)}{\left(e^\beta+1\right)^2}+\frac{e^{2 \beta} \cos ^2\left(\frac{c+\beta}{2}\right) \sin ^2\left(\frac{m+\Delta\theta}{2}\right)}{\left(e^\beta+1\right)^2}\notag\\
&+\frac{e^\beta \cos \left(\frac{c+\Delta\theta}{2}\right) \sin \left(\frac{m+\Delta\theta}{2}\right) \left(-\frac{1}{2} \sin \left(\frac{c}{2}\right) \sin \left(\frac{m}{2}\right) \sin (\Delta\theta)-\sin \left(\frac{c}{2}\right) \cos \left(\frac{m}{2}\right) \sin ^2\left(\frac{\Delta\theta}{2}\right)\right)}{\left(e^\beta+1\right)^2}\notag\\
&+\frac{e^\beta \cos \left(\frac{c+\Delta\theta}{2}\right) \sin \left(\frac{m+\Delta\theta}{2}\right)\cos \left(\frac{c}{2}\right) \cos \left(\frac{\Delta\theta}{2}\right) \sin \left(\frac{m+\Delta\theta}{2}\right)}{\left(e^\beta+1\right)^2},\\
P(2)  &= \frac{e^{2 \beta} \sin ^2\left(\frac{c+\Delta\theta}{2}\right) \sin ^2\left(\frac{m+\Delta\theta}{2}\right)}{\left(e^\beta+1\right)^2}.
\end{align}
With the work probability distribution of work obtained above, we can calculate the correction $Q_{AB}$ in Eq.~\eqref{QABsupp}. From Eqs.~\eqref{Wcumm1} and~\eqref{Wcumm2} in the main text, we get 
\begin{align}\label{W1}
\frac{\langle W^2 \rangle}{N} =&\frac{4 e^{2 \beta} \sin ^2\left(\frac{c+\Delta\theta}{2}\right) \sin ^2\left(\frac{m+\Delta\theta}{2}\right)}{\left(e^\beta+1\right)^2}+\frac{4 \sin ^2\left(\frac{c+\Delta\theta}{2}\right) \sin ^2\left(\frac{m+\Delta\theta}{2}\right)}{\left(e^\beta+1\right)^2}+\frac{e^\beta \sin ^2\left(\frac{c+\Delta\theta}{2}\right) \cos ^2\left(\frac{m+\Delta\theta}{2}\right)}{\left(e^\beta+1\right)^2}\notag\\
&+\frac{e^{2 \beta} \sin ^2\left(\frac{c+\Delta\theta}{2}\right) \cos ^2\left(\frac{m+\Delta\theta}{2}\right)}{\left(e^\beta+1\right)^2}+\frac{\sin ^2\left(\frac{c+\Delta\theta}{2}\right) \cos ^2\left(\frac{m+\Delta\theta}{2}\right)}{\left(e^\beta+1\right)^2}+\frac{e^\beta \cos ^2\left(\frac{c+\Delta\theta}{2}\right) \sin ^2\left(\frac{m+\Delta\theta}{2}\right)}{\left(e^\beta+1\right)^2}\notag\\
&+\frac{e^{2 \beta} \cos ^2\left(\frac{c+\Delta\theta}{2}\right) \sin ^2\left(\frac{m+\Delta\theta}{2}\right)}{\left(e^\beta+1\right)^2}+\frac{\cos ^2\left(\frac{c+\Delta\theta}{2}\right) \sin ^2\left(\frac{m+\Delta\theta}{2}\right)}{\left(e^\beta+1\right)^2}\notag\\
&+\frac{e^\beta \sin \left(\frac{c+\Delta\theta}{2}\right) \cos \left(\frac{m+\Delta\theta}{2}\right) \left(-\frac{1}{2} \sin \left(\frac{c}{2}\right) \sin \left(\frac{m}{2}\right) \sin (\Delta\theta)+\sin \left(\frac{c}{2}\right) \cos \left(\frac{m}{2}\right) \cos ^2\left(\frac{\Delta\theta}{2}\right)\right)}{\left(e^\beta+1\right)^2}\notag\\
&+\frac{e^\beta \sin \left(\frac{c+\Delta\theta}{2}\right) \cos \left(\frac{m+\Delta\theta}{2}\right)\cos \left(\frac{c}{2}\right) \sin \left(\frac{\Delta\theta}{2}\right) \cos \left(\frac{m+\Delta\theta}{2}\right)}{\left(e^\beta+1\right)^2}\notag\\
&+\frac{e^\beta \cos \left(\frac{c+\Delta\theta}{2}\right) \sin \left(\frac{m+\Delta\theta}{2}\right) \left(-\frac{1}{2} \sin \left(\frac{c}{2}\right) \sin \left(\frac{m}{2}\right) \sin (\Delta\theta)-\sin \left(\frac{c}{2}\right) \cos \left(\frac{m}{2}\right) \sin ^2\left(\frac{\Delta\theta}{2}\right)\right)}{\left(e^\beta+1\right)^2}\notag\\
&+\frac{e^\beta \cos \left(\frac{c+\Delta\theta}{2}\right) \sin \left(\frac{m+\Delta\theta}{2}\right)\cos \left(\frac{c}{2}\right) \cos \left(\frac{\Delta\theta}{2}\right) \sin \left(\frac{m+\Delta\theta}{2}\right)}{\left(e^\beta+1\right)^2}
\end{align}

\begin{align}\label{W2}
\frac{\langle W \rangle}{N}  = &\frac{2 e^{2 \beta} \sin ^2\left(\frac{c+\Delta\theta}{2}\right) \sin ^2\left(\frac{m+\Delta\theta}{2}\right)}{\left(e^\beta+1\right)^2}-\frac{2 \sin ^2\left(\frac{c+\Delta\theta}{2}\right) \sin ^2\left(\frac{m+\Delta\theta}{2}\right)}{\left(e^\beta+1\right)^2}+\frac{e^\beta \sin ^2\left(\frac{c+\Delta\theta}{2}\right) \cos ^2\left(\frac{m+\Delta\theta}{2}\right)}{\left(e^\beta+1\right)^2}\notag\\
&+\frac{e^{2 \beta} \sin ^2\left(\frac{c+\Delta\theta}{2}\right) \cos ^2\left(\frac{m+\Delta\theta}{2}\right)}{\left(e^\beta+1\right)^2}-\frac{\sin ^2\left(\frac{c+\Delta\theta}{2}\right) \cos ^2\left(\frac{m+\Delta\theta}{2}\right)}{\left(e^\beta+1\right)^2}+\frac{e^{2 \beta} \cos ^2\left(\frac{c+\Delta\theta}{2}\right) \sin ^2\left(\frac{m+\Delta\theta}{2}\right)}{\left(e^\beta+1\right)^2}\notag\\
&-\frac{e^\beta \cos ^2\left(\frac{c+\Delta\theta}{2}\right) \sin ^2\left(\frac{m+\Delta\theta}{2}\right)}{\left(e^\beta+1\right)^2}-\frac{\cos ^2\left(\frac{c+\Delta\theta}{2}\right) \sin ^2\left(\frac{m+\Delta\theta}{2}\right)}{\left(e^\beta+1\right)^2}\notag\\
&-\frac{e^\beta \sin \left(\frac{c+\Delta\theta}{2}\right) \cos \left(\frac{m+\Delta\theta}{2}\right) \left(-\frac{1}{2} \sin \left(\frac{c}{2}\right) \sin \left(\frac{m}{2}\right) \sin (\Delta\theta)+\sin \left(\frac{c}{2}\right) \cos \left(\frac{m}{2}\right) \cos ^2\left(\frac{\Delta\theta}{2}\right)\right)}{\left(e^\beta+1\right)^2}\notag\\
&-\frac{e^\beta \sin \left(\frac{c+\Delta\theta}{2}\right) \cos \left(\frac{m+\Delta\theta}{2}\right)\cos \left(\frac{c}{2}\right) \sin \left(\frac{\Delta\theta}{2}\right) \cos \left(\frac{m+\Delta\theta}{2}\right)}{\left(e^\beta+1\right)^2}\notag\\
&+\frac{e^\beta \cos \left(\frac{c+\Delta\theta}{2}\right) \sin \left(\frac{m+\Delta\theta}{2}\right) \left(-\frac{1}{2} \sin \left(\frac{c}{2}\right) \sin \left(\frac{m}{2}\right) \sin (\Delta\theta)-\sin \left(\frac{c}{2}\right) \cos \left(\frac{m}{2}\right) \sin ^2\left(\frac{\Delta\theta}{2}\right)\right)}{\left(e^\beta+1\right)^2}\notag\\
&+\frac{e^\beta \cos \left(\frac{c+\Delta\theta}{2}\right) \sin \left(\frac{m+\Delta\theta}{2}\right)\cos \left(\frac{c}{2}\right) \cos \left(\frac{\Delta\theta}{2}\right) \sin \left(\frac{m+\Delta\theta}{2}\right)}{\left(e^\beta+1\right)^2}
\end{align}
\end{widetext}

By expanding the expression of $Q_{AB}$ obtained from Eqs.~\eqref{W1} and~\eqref{W2} to leading order, we obtain
\begin{equation}
    Q_{AB} \approx Nf(\beta)\left[\frac{(c + \Delta \theta)^2}{4} + \frac{(m + \Delta \theta)^2}{4}\right],
\end{equation}
where $f(\beta)$ is given in Eq.~\eqref{f} in the main text.
By setting $c=m=0$, the single-qubit result in Eq.~\eqref{noentangl} is recovered.
In conclusion, we show that an arbitrary separable unitary evolution in the protocol can be understood as a modification of the quench, which can not bring any extra term. In that case, there is only the term proportional to $f(\beta)$ which comes from coherence generation, as discussed in the main text. This shows that the correction term proportional to
$g(\beta)$ only occurs when there is entanglement generation. 

\section{Calculation of the negativity}\label{AppendixC}


For $\hat{R}$ given by Eq.~\eqref{entangler1}, we start by evaluating $\mathcal{N}_0 = \mathcal{N}(\hat{R} \ket{E_0}\bra{E_0}\hat{R}^\dagger)$, where $|E_0\rangle = |00\rangle$, as an example. Notice that,
\begin{widetext}
\begin{equation}
\hat{R}(\ket{E_0}\bra{E_0})\hat{R}^\dagger=\left(
\begin{array}{cccc}
 \cos ^2(c_1-c_2) & 0 & 0 & \frac{\mathrm{i}}{2}   \sin (2 c_1-2 c_2) \\
 0 & 0 & 0 & 0 \\
 0 & 0 & 0 & 0 \\
 -\frac{\mathrm{i}}{2}   \sin (2 c_1-2 c_2) & 0 & 0 & \sin ^2(c_1-c_2) \\
\end{array}
\right).
\end{equation}
In turn, the eigenvalues of the partial transpose of the matrix are given by
\begin{equation}
\begin{aligned}
\left\{-\frac{1}{2} \sqrt{\sin ^2(2 c_1 -2 c_2)},\frac{1}{2} \sqrt{\sin ^2(2 c_1-2 c_2)},\cos ^2(c_1-c_2),\sin ^2(c_1-c_2)\right\}.
\end{aligned}
\end{equation}
\end{widetext}
Taking the absolute value of the negative eigenvalue and expanding the result, for $c_1 - c_2 \ll 1$, gives $\mathcal{N}_0 \approx c_1 - c_2 $. 

More generally, we can calculate $\mathcal{N}(\hat{R}M^{ij}\hat{R}^{\dagger})$ where $i, j$ takes $0$ or $1$, $M^{ij}$ are the projectors defined just after Eq.~\eqref{proj}.
 The results are given by
\begin{widetext}
\begin{align}
&\mathcal{N}(\hat{R}M^{01}\hat{R}^{\dagger})= \mathcal{N}(\hat{R}M^{10}\hat{R}^{\dagger}) = \frac{1}{2} |\sin(2c_1 + 2c_2)|, \\
&\mathcal{N}(\hat{R}M^{00}\hat{R}^{\dagger})= \mathcal{N}(\hat{R}M^{11}\hat{R}^{\dagger}) = \frac{1}{2} |\sin(2c_1 - 2c_2)|.
\end{align}
\end{widetext}
As explained in the main text, we assume that all three parameters, $c_1, c_2, c_3$, are proportional to $N^{-1}$. Therefore, in the slow-driving regime, for $N \gg 1$, we obtain Eqs.~\eqref{negativitydeg} and \eqref{negativitynondeg} in the main text.

\bibliography{bibliography}

\end{document}